\documentstyle{aipproc}
\begin{document}

\title{Conformal gravity and a naturally small cosmological 
constant\footnote{astro-ph/9901219 v2, February 26, 2001.}}
\author{Philip D. Mannheim} 
\address{Department of Physics,
University of Connecticut, Storrs, CT 06269 \\
mannheim@uconnvm.uconn.edu}

\maketitle

\begin{abstract}

With attempts to quench the cosmological constant $\Lambda$ having so far 
failed, we instead investigate what could be done if $\Lambda$ is not 
quenched and actually gets to be as big as elementary particle physics 
suggests. Since the quantity relevant to cosmology is actually
$\Omega_{\Lambda}$, quenching it to its small measured value is equally 
achievable by quenching not $\Lambda$ but $G$ instead, with the $G$ 
relevant to cosmology then being much smaller than that measured in a low 
energy Cavendish experiment. A gravitational model in which this 
explicitly takes place, viz. conformal gravity, is presented, with the 
model being found to provide for a completely natural, non fine tuned 
accounting of the recent high $z$ accelerating universe supernovae 
data, no matter how big $\Lambda$ itself actually gets to be. Thus to 
solve the cosmological constant problem we do not need to change or 
quench the energy content of the universe, but rather only its effect on 
cosmic evolution.
\end{abstract}

The recent discovery \cite{Riess1998,Perlmutter1998} that the current era 
deceleration parameter $q(t_0)$ is close to $-1/2$  has made the already 
extremely disturbing cosmological constant problem even more vexing than 
before. Specifically, with $q(t_0)$ being given in standard gravity by 
$q(t_0)=(n/2-1)\Omega_{M}(t_0)-\Omega_{\Lambda}(t_0)$ [where 
$\Omega_{M}(t)=8\pi G\rho_{M}(t)/3c^2H^2(t)$ is due to ordinary matter 
(i.e. matter for which $\rho_{M}(t)=A/R^n(t)$ where $A>0$ and 
$3 \leq n \leq 4$), and where  
$\Omega_{\Lambda}(t)=8\pi G\Lambda/3cH^2(t)$ is due to a cosmological 
constant], we see that not only must $c\Lambda$ be non-zero, it must be 
of order $3c^2H^2(t_0)/8\pi G=\rho_{C}(t_0)$ in magnitude, i.e. it must be 
quenched by no less than 60 orders of magnitude below its natural value as 
expected from fundamental particle physics. Additionally, since such a 
quenched $c\Lambda$ would then be of order $\rho_{M}(t_0)$ as well (the 
so-called cosmic coincidence), our particular cosmological epoch would 
then only be achievable in standard gravity if the macroscopic Friedmann 
evolution equation were to be fine-tuned at very early times to 
incredible precision. Any still to be found fundamental microscopic 
physics mechanism which might in fact quench $c\Lambda$ by the requisite 
sixty orders of magnitude would thus still leave standard gravity with an 
additional macroscopic coincidence to explain. 

Since no mechanism has yet been found which might actually quench 
$c\Lambda$ and since its quenching might not necessarily work 
macroscopically anyway, we shall thus turn the problem around and ask what 
can be done if $c\Lambda$ is not in fact quenched and is in fact as big as 
elementary particle physics suggests. To this end we note immediately that 
it would still be possible to have $q(t_0)$ be of order one today (the 
measurable consequence of $c\Lambda$) if instead of quenching $c\Lambda$ 
we instead quench $G$, with the cosmological $G$ then being replaced by an 
altogether smaller $G_{eff}$. Since observationally $\rho_{M}(t_0)$ is 
known to not be bigger than $\rho_{C}(t_0)$, any successful such
cosmological quenching of $G$ (successful in the sense that such 
relativistic quenching not modify standard non-relativistic physics) 
would immediately leave us with a non-quenched $c\Lambda$ which would 
then not suffer from any cosmic coincidence problem. 

Given these remarks it is thus of interest to note that it is precisely a 
situation such as this which obtains in the conformal gravity theory which 
has recently been advanced 
\cite{M1990,M1992,M1994,M1997,M1998,M1999,M2000} as 
a candidate alternative to the standard gravitational theory. Conformal 
gravity is a fully covariant gravitational theory which, unlike standard 
gravity, possesses an additional local scale invariance, a symmetry which 
when unbroken sets any fundamental cosmological constant and any 
fundamental $G$ to zero\cite{M1990}. Unlike standard gravity conformal 
gravity thus has a great deal of control over the cosmological constant, a 
control which is found to be of relevance even after the conformal 
symmetry is spontaneously broken by the non-vanishing of a scalar field 
vacuum expectation value $S_0$ below a typical critical temperature 
$T_V$. In fact in the presence of such breaking the standard attractive 
$G$ phenomenology is found to still emerge at low energies \cite{M1994}, 
while cosmology is found \cite{M1992} to instead be controlled by the 
effective $G_{eff}=-3c^3/4\pi \hbar S_0^2$, a quantity which by being 
negative immediately entails cosmic repulsion \cite{M1998}, and which, 
due to its behaving as $1/S_0^2$, is made small by the very same 
mechanism which serves to make $\Lambda$ itself large. 

Other than the use of a changed $G$ the cosmic evolution of conformal 
gravity is otherwise the same as that of the standard one, viz. 
\cite{M1998,M1999,M2000} 
\begin{eqnarray}
\dot{R}^2(t) +kc^2 =
-3c^3\dot{R}^2(t)(\Omega_{M}(t)+
\Omega_{\Lambda}(t))/ 4 \pi \hbar S_0^2 G 
\equiv \dot{R}^2(t)(\bar{\Omega}_{M}(t)+
\bar{\Omega}_{\Lambda}(t)) 
\nonumber \\
q(t)= (n/2-1)\bar{\Omega}_{M}(t)-\bar{\Omega}_{\Lambda}(t)
\label{1}
\end{eqnarray}
(Eq. (\ref{1}) serves to define $\bar{\Omega}_{M}(t)$ and
$\bar{\Omega}_{\Lambda}(t)$). Moreover, unlike the situation in the 
standard theory where values for the relevant evolution parameters (such 
as the sign of $\Lambda$) are only determined phenomenologically, in 
conformal gravity essentially everything is already a priori known. With 
conformal gravity not needing dark matter to account for 
non-relativistic issues such as galactic rotation curve systematics 
\cite{M1997}, $\rho_{M}(t_0)$ can be determined directly from luminous 
matter alone, with galaxy luminosity accounts giving a value for it of 
order $0.01\rho_C(t_0)$ or so. Further, with $c\Lambda$ being generated 
by vacuum breaking in an otherwise scaleless theory, since such breaking 
lowers the energy density, $c\Lambda$ is unambiguously negative, with it
thus being typically given by $-\sigma T_V^4$. Then with $G_{eff}$ also 
being negative, $\bar{\Omega}_{\Lambda}(t)$ is necessarily 
positive, just as needed to give cosmic acceleration. 
Similarly, the sign of the spatial 3-curvature $k$ is known from theory 
\cite{M2000} to be negative,\footnote{At the highest temperatures the 
zero energy density required of a (then) completely conformal invariant 
universe is maintained by a cancellation between the positive energy 
density of ordinary matter and the negative energy density 
due to the negative curvature of the gravitational field.} something 
which has been independently confirmed from a study of 
galactic rotation curves \cite{M1997}. Finally, since $G_{eff}$ is 
negative, the cosmology is singularity free and thus expands from a 
finite maximum temperature $T_{max}$, a temperature which for 
$k<0$ is necessarily greater than $T_V$ \cite{M1998,M1999,M2000} (so that 
a large $T_V$ entails an even larger $T_{max}$). 

Given only that $\Lambda$, $k$ and $G_{eff}$ are all negative, the 
temperature evolution of the theory is then completely determined
for arbitrary $T_{max}$ and $T_V$, to yield \cite{M1998,M1999,M2000} 

\begin{equation}
\bar{\Omega}_{\Lambda}(t)= 
(1-T^2/T_{max}^2)^{-1}(1+T^2T_{max}^2/T_V^4)^{-1},~~ 
\bar{\Omega}_M(t)=-(T^4/T_V^4)\bar{\Omega}_{\Lambda}(t)
\label{2}
\end{equation}
at any $T$.
Thus, from Eq. (\ref{2}) we see that simply because $T_{max} \gg T(t_0)$, 
i.e. simply because the universe is as old as it is, it immediately 
follows that $\bar{\Omega}_{\Lambda}(t_0)$ has to lie somewhere between 
zero and one today no matter how big (or small) $T_V$ might be. Then, 
since $T_V \gg T(t_0)$, $\bar{\Omega}_M(t_0)$ has to be completely 
negligible,\footnote{$\bar{\Omega}_M(t_0)$ is suppressed by 
$G_{eff}$ being small, and not by $\rho_{M}(t_0)$ itself being small.} so 
that $q(t_0)$ must thus necessarily lie between zero and minus one today 
notwithstanding that $T_V$ is huge. Moreover, the larger $T_V$ gets  
to be, the more $\bar{\Omega}_{\Lambda}(t_0)$ will be reduced below one, 
with it taking a value close to one half should $T(t_0)T_{max}/T_V^2$ be 
close to one. With $\bar{\Omega}_M(t_0)$ being 
negligible today, $\bar{\Omega}_{\Lambda}(t_0)$ is therefore given as 
$1+kc^2/\dot{R}^2(t_0)$, a quantity which necessarily lies below one if 
$k$ is negative. Thus in a $k<0$ conformal gravity universe, once 
the universe has cooled enough, $\bar{\Omega}_{\Lambda}(t)$ will then be 
forced to have to lie between zero and one no matter how big $\Lambda$ 
may or may not be. The contribution of $\Lambda$ to cosmology is thus 
seen to be completely under control in conformal gravity, with the theory 
thus leading us right into the $\bar{\Omega}_{\Lambda}(t_0) \simeq 1/2$, 
$\bar{\Omega}_M(t_0)=0$ region, a region which, while foreign to 
standard gravity, is nonetheless still fully compatible with the reported 
supernovae data fits. Hence to solve the cosmological constant problem we 
do not need to change or quench the energy content of the universe, but 
rather only its effect on cosmic evolution.
This work has been supported in part by the 
Department of Energy under grant No. DE-FG02-92ER40716.00.

\end{document}